\pdfoutput=1

\documentclass[sigconf,table]{acmart}

\renewcommand\footnotetextcopyrightpermission[1]{} 
\usepackage{amsmath}
\usepackage{graphicx}
\usepackage{booktabs}
\usepackage{amssymb}
\usepackage{latexsym}	
\usepackage{array}		
\usepackage{multirow}
\usepackage{subfigure}
\usepackage{algorithm}
\usepackage[noend]{algpseudocode}
\usepackage{threeparttable}
\usepackage{paralist}   
\usepackage{xspace}
\usepackage{color}
\usepackage{adjustbox}
\usepackage{balance}
\usepackage{wasysym}
\usepackage{rotating}   
\usepackage{listings}
\usepackage{flushend}  
\usepackage{tabu}
\usepackage{amssymb}
\usepackage{pifont}

\usepackage{url}            

\usepackage{xpatch}
\makeatletter
\chardef\TPT@@@asteriskcatcode=\catcode`*
\catcode`*=11
\xpatchcmd{\threeparttable}
  {\TPT@hookin{tabular}}
  {\TPT@hookin{tabular}\TPT@hookin{tabu}}
  {}{}
\catcode`*=\TPT@@@asteriskcatcode
\makeatother

\setcopyright{none}

\newcommand{\red}[1]{\textcolor[rgb]{1.00,0.00,0.00}{#1}}
\newcommand{\blue}[1]{\textcolor[rgb]{0.00,0.00,1.00}{#1}}
\newcommand{\darkblue}[1]{\textcolor[rgb]{0.00,0.00,0.65}{#1}}
\newcommand{\green}[1]{\textcolor[rgb]{0.00,0.60,0.00}{#1}}

\newcommand{\cha}{\red{\ding{55}}}
\newcommand{\gou}{\green{\ding{52}}}
\newcommand{\ling}{\darkblue{\RIGHTcircle}}

\definecolor{wheat1}{rgb}{1.000000,0.905882,0.729412}
\definecolor{LightGray}{rgb}{0.827451,0.827451,0.827451}

\newcolumntype{a}{>{\columncolor{wheat1}}l}

\definecolor{mygreen}{rgb}{0,0.6,0}
\definecolor{mygray}{rgb}{0.5,0.5,0.5}
\definecolor{mymauve}{rgb}{0.58,0,0.82}
\definecolor{darkblue}{rgb}{0.0,0.0,0.6}
\definecolor{maroon}{RGB}{102, 0, 0}
\definecolor{Maroon}{cmyk}{0,0.87,0.68,0.32}
\definecolor{darkred}{RGB}{139, 0, 0}
\definecolor{forestgreen}{RGB}{34, 139, 34}

\lstdefinelanguage{XML}
{
  basicstyle=\ttfamily\small,   
  morestring=[b]",
  moredelim=[s][\color{darkblue}]{<}{\ },
  moredelim=[s][\color{darkblue}]{</}{>},
  moredelim=[l][\color{darkblue}]{/>},
  moredelim=[l][\color{darkblue}]{>},
  morecomment=[s]{<?}{?>},
  morecomment=[s]{<!--}{-->},
  stringstyle=\color{darkred},
  identifierstyle=\color{mymauve}
}

\lstdefinestyle{customJava}{
  breaklines=true,
  keepspaces=true,
  frame=single,
  language=Java,
  showstringspaces=false,
  basicstyle=\footnotesize\ttfamily,
  keywordstyle=\color{blue},
  otherkeywords={+, getIntent},
  numbers=left,
  numbersep=5pt,
  numberstyle=\scriptsize\color{black},
  rulecolor=\color{black},
  stepnumber=1,
  tabsize=2,
  commentstyle=\itshape\color{green!40!black},
  stringstyle=\color{orange},
  emph=[1]  
  {
        do,
        try,
        new,
        catch,
        while,
        SecProvider,
        SecReceiver,
        SecService,
        SecActivity,
        SecSink,
  },
  emphstyle=[1]{\color{darkred}},
  emph=[2]  
  {
        @Override,
  },
  emphstyle=[2]{\color{purple!40!black}},
  belowskip=-1em, 
}

\newif\ifANNOYMIZE
\ANNOYMIZEfalse

\newif\ifACM
\ACMtrue  

\ifACM
\fi

\ifACM
\newcommand{\myfig}{Figure\xspace}
\else
\newcommand{\myfig}{Fig.\xspace}
\fi

\ifACM
\newcommand{\mysec}{\S}
\else
\newcommand{\mysec}{Section\xspace}
\fi

\newcommand{\name}{SCLib\xspace}  

\newcommand{\activity}{Activity\xspace}
\newcommand{\service}{Service\xspace}
\newcommand{\receiver}{Receiver\xspace}
\newcommand{\provider}{Provider\xspace}
\newcommand{\intent}{Intent\xspace}
\newcommand{\binder}{Binder\xspace}

\begin{document}

\title{Where is the Road? Our Vision on Practically Defending Against Component Hijacking in Android Applications}
\title{No More Hijack: Our Vision on Practically Defending Against Component Hijacking in Android Applications}
\title{Our Vision on Practically Defending Against Component Hijacking in Android Applications}
\title{\name: Towards Practically Defending Against Component Hijacking in Android Applications}
\title{\name: A Practical and Lightweight Approach to Defending Against Component Hijacking in Android Applications}
\title{\name: A Practical and Lightweight Defense against Component Hijacking in Android Applications}
\subtitle{This is the extended technical report version of our SCLib paper in ACM CODASPY 2018~\cite{SCLib18}.}

\author{Daoyuan Wu}
\affiliation{\institution{School of Information Systems, Singapore Management University}}
\email{dywu.2015@smu.edu.sg}

\author{Yao Cheng}
\authornote{Work by this author was performed while at Singapore Management University.}
\affiliation{\institution{Institute for Infocomm Research, A*STAR, Singapore}}
\email{cheng\_yao@i2r.a-star.edu.sg}

\author{Debin Gao, Yingjiu Li, and Robert H. Deng}
\affiliation{\institution{School of Information Systems, Singapore Management University}}
\email{{dbgao, yjli, robertdeng}@smu.edu.sg}

\renewcommand{\shortauthors}{D. Wu et al.}

\begin{abstract}

Cross-app collaboration via inter-component communication is a fundamental mechanism on Android.
Although it brings the benefits such as functionality reuse and data sharing, a threat called \textit{component hijacking} is also introduced.
By hijacking a vulnerable component in victim apps, an attack app can escalate its privilege for operations originally prohibited.
Many prior studies have been performed to understand and mitigate this issue, but \textit{no} defense is being deployed in the wild, largely due to the deployment difficulties and performance concerns.
In this paper we present \name, a \textit{secure component library} that performs in-app mandatory access control on behalf of app components.
It does not require firmware modification or app repackaging as in previous works.
The library-based nature also makes \name more accessible to app developers, and enables them produce secure components in the first place over fragmented Android devices.
As a proof of concept, we design six mandatory policies and overcome unique implementation challenges to mitigate attacks originated from both system weaknesses and common developer mistakes.
Our evaluation using ten high-profile open source apps shows that \name can protect their 35 risky components with negligible code footprint (less than 0.3\% stub code) and nearly no slowdown to normal intra-app communications.
The worst-case performance overhead to stop attacks is about 5\%.

\end{abstract}

\maketitle

\section{Introduction}
\label{sec:intro}

Android has been the dominant player in smartphone markets in the last few years.
On Android, different apps collaborate with each other via inter-component communication~\cite{UnderstandAndroid09}.
Although such flexible cross-app collaboration brings the benefits of functionality reuse and data sharing, \textit{component hijacking}~\cite{CHEX12} is also introduced in which an attack app hijacks a vulnerable component in victim apps to bypass Android sandbox and escalate its privilege~\cite{ISC10_Privilege}, causing confused deputy problems~\cite{Confused88} such as permission misuse~\cite{Woodpecker12}, data manipulation~\cite{CHEX12}, and content leaks~\cite{ContentScope13}.

Many approaches have been proposed to mitigate component hijacking.
One major line of the research~\cite{Saint09, IPCInspection11, Quire11, TrustDroid11, Taming12, IntraComDroid12, Scippa14, IEM16} is to modify and extend the Android operating system to supervise inter-component communication.
The other direction~\cite{AppSealer14} is to patch app binaries with repackaging~\cite{Aurasium12}.
Both are useful if they could be deployed in the wild, but nearly no proposal\footnote{\small Only SEAndroid~\cite{SEAndroid13} was adopted to replace the Linux UID-based sandbox with the SELinux-confined sandbox~\cite{SELinuxAndroid}.} has been integrated into Android or adopted by Google Play to date, largely due to the compatibility and performance concerns.
For example, repackaging violates Android's app verification mechanism and thus is not favorable by app markets and developers who own the source code.
Consequently, component hijacking remains a serious open problem in the Android ecosystem.
As one of our contributions, we make a comprehensive comparison on existing defenses in \mysec\ref{sec:related}.

\begin{figure}[h!]
\vspace{-3ex}
\centering
  \subfigure[\small Developer-driven deployment via regular app updates.] {
    \includegraphics[width=0.39\textwidth]{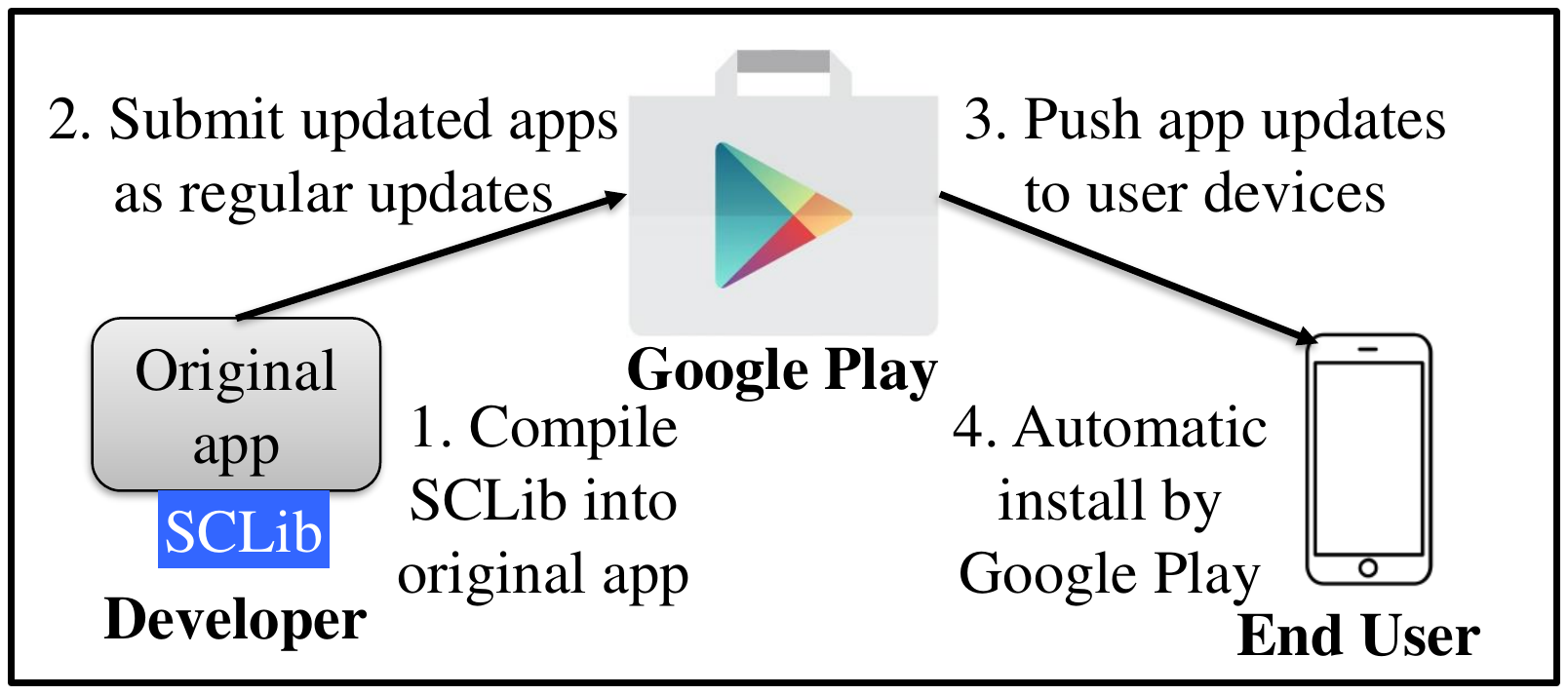}
    \label{fig:deploymodel1}
  }
  \subfigure[\small End user-driven deployment via Boxify~\cite{Boxify15}.] {
    \includegraphics[width=0.39\textwidth]{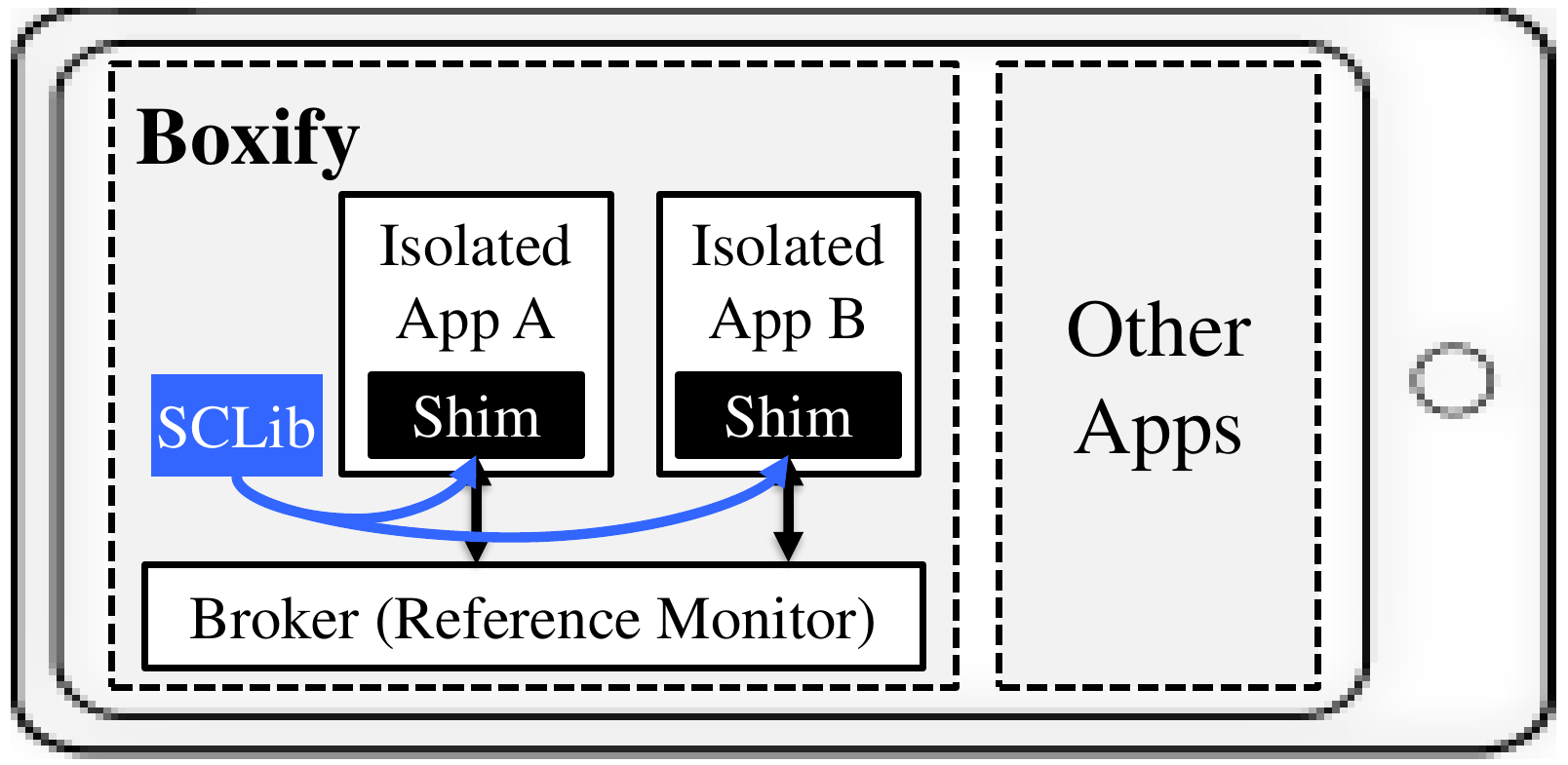}
    \label{fig:deploymodel2}
  }
\vspace{-3ex}
\caption{Two deployment models of \name.}
\vspace{-2ex}
\label{fig:deployment}
\end{figure}

\textbf{Key idea.}
In this paper, we provide a new perspective to practically defending against component hijacking.
Our solution is a \textit{secure component library}, shorted as \name, which performs in-app mandatory access control on behalf of app components.
Due to its library-based nature, \name requires neither firmware modification nor app repackaging, significantly reducing the deployment difficulties.
Specifically, we propose two deployment models as shown in \myfig~\ref{fig:deployment}, the developer-driven and the end user-driven deployment.

\textbf{Deployment models.}
\name can be compiled by app developers into their original apps in their regular app updates (e.g., for functionality improvement), which are then pushed to user devices and automatically installed by Google Play.
This deployment model introduces minimal burden to developers
because they have already accumulated experiences to integrate third-party libraries such as OkHttp~\cite{OkHttp} and advertisement libraries.
Further, \name can help developers secure their apps in the first place (rather than applying patches after apps have been released) over fragmented Android devices~\cite{Fragmentation} (a major limitation of firmware modification approaches).

To further enable end users to secure their apps directly, we envision the second deployment model through state-of-the-art app sandboxing technology, e.g., Boxify, which sandboxes any other app into its own process space and delegates their inter-component communication via a reference monitor called Broker.
Note that Boxify does not require root privilege, firmware modification, or app repackaging.
We refer interested readers to~\cite{Boxify15} for more details.
As shown in \myfig~\ref{fig:deploymodel2}, \name can be plugged into Boxify as part of its policy module or its shim code in each isolated app.
Since \name's design is generally the same for both deployment models, we present it under the first deployment in the rest of this paper.

\textbf{SCLib design.}
As a major component of \name, we devise a set of practical in-app policies to defend against component hijacking.
Our policy checking is based on enforcement primitives that previous efforts have not fully leveraged, including various component attributes and input data of incoming requests.
\name automatically collects these primitives at entry points of the protected components, and enforces ``just-enough'' policies from the pre-defined policy set.
As a proof of concept, we design six mandatory policies that either directly deny illegal requests or alert users via a pop-up dialog for suspicious requests.
These policies can mitigate component hijacking originated from both system weaknesses and common developer mistakes, half of which have not been tackled by previous efforts.
Moreover, we design \name to cover all four types of components for the first time (see \mysec\ref{sec:related}).

In the course of implementing \name, we identify and overcome three major challenges that are unique in our context.
First, Android currently fails to provide the caller identity information to most callee components, as explained in~\cite{Scippa14}.
This caller identity, however, is essential in implementing our mandatory access control policies.
Unlike the previous solution~\cite{Scippa14} that modifies Android source code, \name leverages the Binder side channel to recover caller app identities at the application layer (see \mysec\ref{sec:implementCaller}).
Second, it is nearly infeasible to pop up alert dialog in the intercepted components due to the lack of appropriate user interface context and limited function return timing thresholds.
We solve this problem by a novel dialog-like Activity transition technique, which overcomes both context and timing restrictions while maintaining user experience and policy enforcement logic (see \mysec\ref{sec:implementUI}).
Third, there is lack of API support to collect certain component attributes (e.g., whether an exported component is explicitly or implicitly exported).
\name performs runtime Android manifest analysis by itself (see \mysec\ref{sec:implementAttr}).

\name is a lightweight solution by design.
It enforces policy checking only at the entry points, and thus has no additional overhead of information flow tracking that is required in some existing approaches, e.g., AppSealer~\cite{AppSealer14}.
In addition, \name only affects the performance of \textit{certain} exported components that require protection.
In contrast, hooking-based checking, e.g., Aurasium~\cite{Aurasium12}, adds overhead to both exported and non-exported components.
Firmware modification approaches introduce overhead to all inter- and intra-app communications in all apps within the system.

\textbf{Evaluation.}
We evaluate \name using ten high-profile open source apps.
We show that these well-tested apps contain 35 risky components that \name can contribute more protection.
Our measurement further finds that \name introduces negligible code footprint --- less than 0.3\% stub code in all cases.
Furthermore, by performing eight detailed security case studies, we demonstrate \name's unique values as compared to developers' own patches and Android platform updates.
Finally, our performance evaluation shows that \name incurs modest overhead to those protected components (no overhead at all to other components).


The remainder of this paper is as follows.
We first introduce the threat model in \mysec\ref{sec:model}, outline the objectives and analyze existing solutions in \mysec\ref{sec:goalANDrelated}.
The design and implementation of \name are presented in \mysec\ref{sec:seclib}.
In \mysec\ref{sec:evaluate}, we evaluate \name's efficacy and efficiency, followed by a discussion in \mysec\ref{sec:discuss}.
Finally, we conclude the paper in \mysec\ref{sec:conclude}.

\section{Threat Model}
\label{sec:model}

\begin{figure}[t!]
\begin{adjustbox}{center}
\includegraphics[width=0.3\textwidth]{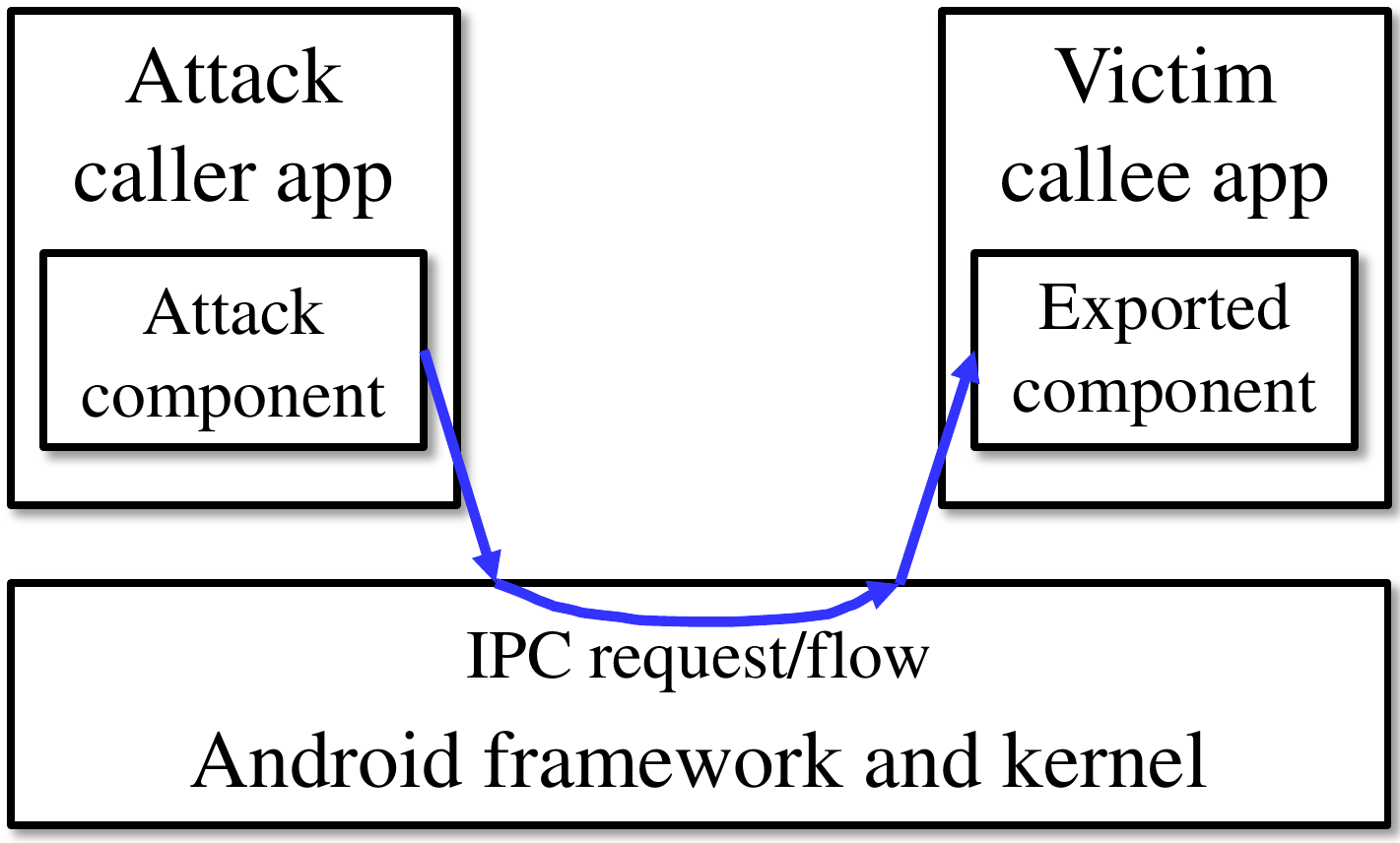}
\end{adjustbox}
\vspace{-5ex}
\caption{The threat model of component hijacking.}
\label{fig:threatmodel}
\vspace{-4ex}
\end{figure}

\myfig~\ref{fig:threatmodel} presents our threat model of component hijacking on Android.
The adversary is a \textit{caller app}, and the victim is a \textit{callee app} that contains a component that is \textit{exported}.
The attack component in the caller app sends a crafted \textit{IPC} (inter-process communication)\footnote{\small A.k.a. ICC (inter-component communication)~\cite{ICCEpicc13}.} request to the exported component to maliciously trigger its code execution for a privileged operation, e.g., permission misuse~\cite{Woodpecker12} and data manipulation~\cite{CHEX12}.
In this sense, component hijacking belongs to the classic confused deputy problem~\cite{Confused88}.
Note that although \myfig~\ref{fig:threatmodel} shows only two parties, our defense can handle hijacking via one or multiple middle app(s).

More specifically, we underline two in-scope threats that are not considered in some related works. 
\begin{compactitem}
\item Unlike some existing work~\cite{Woodpecker12, ContentScope13}, we do \textit{not} assume that exported components protected with above-\texttt{normal} permissions~\cite{PermissionLevel} are always safe.
We do not consider it safe because for an exported component protected with a \texttt{dangerous}-level permission, an attack app can still register the corresponding permission for sending IPC requests.
Additionally, a recent report~\cite{CustomPermission} showed that even components with a \texttt{signature}-level permission could be compromised, because the attack app can pre-claim that permission as \texttt{normal} if it is installed earlier than the victim app.

\item Similarly, for the attack app, we do \textit{not} assume that it always has zero or few permissions since it can claim the same permission as the misused permission in a victim app.
The benefit for doing so is that it may deceive the IPC call chain-based permission checking~\cite{IPCInspection11, Quire11}.
We \textit{do} assume that the attack app has no root privilege though.

\end{compactitem}

It is worth noting that a related threat called \textit{unauthorized Intent receipt}~\cite{ComDroid11} is out of the scope of this paper.
This threat is essentially different from component hijacking.
In its model, the attack app is the callee, and exploit occurs because the caller victim app mistakenly sends out sensitive information in its IPC messages.
A systematic defense~\cite{Aquifer13} has been proposed to mitigate this issue.

Additionally, we do \textit{not} consider that the host app would break the integrity of \name or its policies.
This is reasonable because our objective is to protect potentially vulnerable apps instead of the malicious apps.
Similarly, the malicious app colluding~\cite{Collusion12} is also out of our scope.

\section{Objectives and Related Works}
\label{sec:goalANDrelated}

\subsection{Design Objectives}
\label{sec:objective}

To defend against component hijacking in the wild, we identify the following four objectives: 

\begin{compactitem}
\item[O1]
\textbf{No firmware modification.}
The approach does not rely on firmware customization. 
It should also work without the root privilege.

\item[O2]
\textbf{No app repackaging.}
The approach does not repackage target apps in bytecode or binary rewriting~\cite{DroidMOSS12, PushPopRandom15}.

\item[O3]
\textbf{Handling all four types of components.}
The proposed solution shall protect all four types of Android components, including Intent-based components (i.e., \activity, \service, and \receiver) and non-Intent based components (i.e., \provider). Note that in this paper, we simplify BroadcastReceiver and ContentProvider as Receiver and Provider, respectively.

\item[O4]
\textbf{Minimal impact on normal operations.}
The solution has minimal performance impact on normal app functionality and intra-app communication.
\end{compactitem}


\subsection{Analysis of Existing Solutions}
\label{sec:related}

We now analyze how existing solutions defend against component hijacking and the extent to which they achieve the aforementioned four objectives.
Table~\ref{tab:related} summarizes our analysis on major defenses against component hijacking.

First, most existing defenses require firmware modification (O1: \cha).
This includes Saint~\cite{Saint09} for adding install- and run-time policies that require developers to specify,
IPC Inspection~\cite{IPCInspection11}, Quire~\cite{Quire11}, TrustDroid~\cite{TrustDroid11}, Scippa~\cite{Scippa14}, Bugiel et al.~\cite{Taming12} for using system-wide reference monitors to check inter-component call chains to prevent privilege (mainly permissions~\cite{Woodpecker12}) escalation,
and Kantola et al.~\cite{IntraComDroid12} for inferring and restricting unintentional component exposure.
More recently, IEM~\cite{IEM16} extends the Android framework to enable user-layer Intent firewall apps.
There are other related approaches (e.g.,~\cite{FlaskDroid13, ASM14, ASF14, CICC15, DIFC16}) in this category, though they were not specialized for defending against component hijacking.
All these prior works take advantage of the open-sourced nature of Android to make code changes and provide more secure OS design principles, but in the real world, they were not adopted by smartphone vendors.
Additionally, end users have no capability to flash the modified firmware in general.

\begin{table}[t!]
\caption{A comparison of major defenses against component hijacking.}
\vspace{-3ex}
\label{tab:related}
\begin{adjustbox}{center}
\scalebox{0.8}{
\begin{threeparttable}
\begin{tabular}{|l|l|c|c|c|c|}

\cline{2-6}
\multicolumn{1}{c|}{} & \multicolumn{1}{c|}{\multirow{2}{*}{Core Idea}} & \multicolumn{4}{c|}{Objectives (see \mysec\ref{sec:objective})} \\
\cline{3-6}
\multicolumn{1}{c|}{} & & O1 & O2 & O3 & O4 \\
\hline

Saint~\cite{Saint09} & Adding install- and run-time policies & \cha & \gou & \ling & \cha \\
\hline

IPC Inspection~\cite{IPCInspection11} & Checking IPC call chains & \cha & \gou & \cha & \cha \\
\hline

Quire~\cite{Quire11} & Checking IPC and RPC call chains & \cha & \gou & \cha & \cha \\
\hline

TrustDroid~\cite{TrustDroid11} & Mediating IPC in middleware layer& \cha & \gou & \ling & \cha \\
\hline

Bugiel et al.~\cite{Taming12} & Mediating IPC in different layers& \cha & \gou & \ling & \cha \\
\hline

Aurasium~\cite{Aurasium12} & Intercepting sensitive API calls& \gou & \cha & \ling & \ling\\
\hline

Kantola et al.~\cite{IntraComDroid12} & Restricting component exposure & \cha & \gou & \cha & \gou \\
\hline

IntentFirewall~\cite{IntentFirewallCode} & A system-layer firewall to check Intents & \ling & \gou & \cha & \cha \\
\hline

AppSealer~\cite{AppSealer14} & Flow checking of incoming Intents & \gou & \cha & \cha & \cha \\
\hline

Scippa~\cite{Scippa14} & Building system-centric IPC call chains & \cha & \gou & \cha & \cha \\
\hline

IEM~\cite{IEM16} & Enabling user-layer Intent firewall apps & \cha & \gou & \cha & \cha \\
\hline

\end{tabular}
\begin{tablenotes}
\item\centering \gou = applies; \ling = partially applies; \cha = does not apply.
\end{tablenotes}
\end{threeparttable}
}
\end{adjustbox}
\end{table}

A particularly interesting example is IntentFirewall~\cite{IntentFirewallCode}.
Although it was introduced into the Android Open Source Project (AOSP) repository over four years ago, it is still experimental and not an officially supported feature of the Android framework~\cite{IntentFirewallDoc}, probably due to its limitations~\cite{IntentFirewallDoc}.
The SEAndroid community is exploring the idea of using IntentFirewall as a potential replacement of their experimental Intent MAC mechanism~\cite{IntentFirewallHomepage}, because in the original form of SEAndroid~\cite{SEAndroid13}, it does not audit app-layer IPC. 
SEAndroid tries to reconstruct Android's sandbox from the previous Linux UID-based discretionary access control to the present SELinux-confined mandatory access control.
It can restrict certain app flaws such as direct file leak~\cite{SEAndroid13} but not component hijacking or indirect file leak~\cite{MoST15, FileCross14}, because it is challenging to efficiently audit every IPC at the system level without affecting normal app functionality. 
Even if one day a solid Intent MAC might be activated, it still faces the deployment challenges to protect fragmented and outdated devices.

Second, other defenses typically need to perform app repackaging (O2:~\cha).  
Aurasium~\cite{Aurasium12} and AppSealer~\cite{AppSealer14} are the two notable examples.
Specifically, Aurasium repackages apps to insert API hooking code to intercept sensitive API calls.
It mainly aims to prevent malware but can also be used to mitigate component hijacking in which sensitive APIs (in a victim app) are triggered by an attack app.
On the other hand, AppSealer is specialized to generate patched apps that introduce flow tracking to avoid critical APIs being triggered by malicious Intents.
Both approaches are attractive from the security's perspective, but they also face the deployment difficulties:
(i) repackaging is unlikely adopted by app developers because they own the source code and do not want repackaging to affect the original code quality;
(ii) app stores are unlikely to deploy repackaging-based approaches because they (including Google Play) have no access to the developers' private signing keys.

Third, none of the prior efforts has fully handled all four types of components (O3:~\cha; O3: \ling).
Most of them only protect the Intent-based components, whereas the non-Intent based \provider component is largely under-treated.
Although all underlying Android IPC communications go through the Binder driver~\cite{BinderIntro, Binder11}, Intent and \provider are two different higher-level abstractions of Binder~\cite{DiveBinder13}.
Existing approaches either just modify the Android framework to supervise Intent IPC (e.g.,~\cite{IPCInspection11, Quire11}), or only recover Intents' semantic from the kernel-layer Binder (e.g.,~\cite{Taming12, Scippa14}).
Approaches such as AppSealer~\cite{AppSealer14} and IEM~\cite{IEM16} also explicitly target at Intent-based components.
Note that although some approaches (e.g., \cite{Saint09, TrustDroid11}) mentioned the protection for \provider to some extent, the generic and broader app \provider vulnerabilities~\cite{ContentScope13} were not touched because the problem itself was discovered only afterwards.
Probably for the same reason, existing solutions focus only on permission misuse and data manipulation, yet the content leak and pollution problem in \provider has to rely on Android's updates.

Fourth, nearly no defenses satisfy the requirement of minimal checking on normal operations (O4:~\cha).
Except the work from Kantola et al.~\cite{IntraComDroid12}, firmware modification approaches have to monitor all IPC communications in all apps within the system.
Although they achieve the whole-system coverage, the performance was sacrificed by not focusing on apps or components that need protection.
Moreover, they often need to retrieve the corresponding permission for each call chain, which also increases the overhead.
On the other hand, Intent flow based repackaging approaches such as AppSealer~\cite{AppSealer14} can concentrate on protecting risky app components, but its data flow tracking is expensive.
Hooking-based API checking in Aurasium~\cite{Aurasium12} is lightweight; however, it does not differentiate sensitive API calls resulting from user operations or malicious IPC.

Lastly, as orthogonal to the defense research, many prior studies try to understand and detect component hijacking issues in real-world apps.
They have leveraged program analysis techniques to propose various detection mechanisms, including ComDroid~\cite{ComDroid11}, Woodpecker~\cite{Woodpecker12}, CHEX~\cite{CHEX12}, DroidChecker~\cite{DroidChecker12}, ContentScope~\cite{ContentScope13}, Epicc~\cite{ICCEpicc13}, ECVDetector~\cite{ECVDetector14}, WeChecker~\cite{WeChecker15}, and DIALDroid~\cite{DIALDroid17}.
Android app analysis frameworks, such as FlowDroid~\cite{FlowDroid14} and Amandroid~\cite{Amandroid14}, could also be extended to analyze component hijacking.
Recently, more solid and scalable inter-component analysis methods~\cite{IC315, MarketICC16} and an Intent policy checking system called IntentScope~\cite{IntentScope16} are proposed.

\section{\name: Secure Component Library}
\label{sec:seclib}

This section covers the design and implementation details of \name.
We begin with an overview of \name and some challenges in its design
in \mysec\ref{sec:overview}, and then present some important MAC
policies that \name is capable of enforcing
(\mysec\ref{sec:policy}).  After that, we discuss the detailed
implementation of \name with focus on our novel ways of handling the
challenges.

\subsection{Design Overview}
\label{sec:overview}

In this subsection, we first show a big picture of \name, and then highlight the major challenges in designing and implementing \name.

\begin{figure}[t!]
\vspace{-1ex}
\begin{adjustbox}{center}
\includegraphics[width=0.4\textwidth]{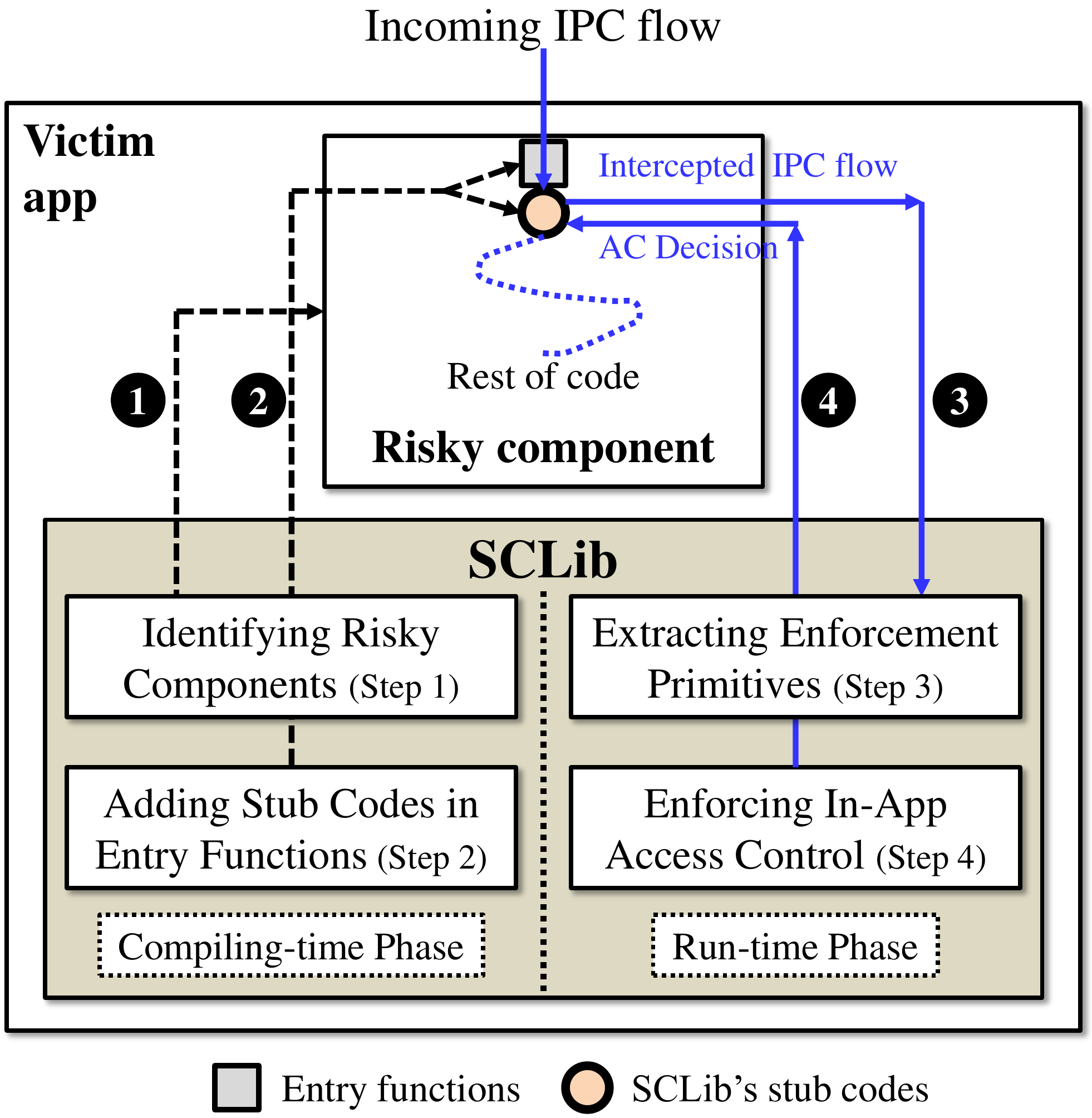}
\end{adjustbox}
\vspace{-5ex}
\caption{A high-level overview of \name.}
\label{fig:seclib}
\vspace{-4ex}
\end{figure}

\myfig~\ref{fig:seclib} presents the overall design of \name.
At the high-level view, \name is a regular user-space library that could be easily integrated into apps on different Android platforms.
\name aims to be a secure component library that performs in-app mandatory access control (MAC) on behalf of app components to defend against component hijacking.
With a set of pre-defined MAC policies in \name, developers can overcome the by-default system weaknesses and common mistakes.
Moreover, due to the library-based nature, \name inherently requires \textit{no} firmware modification or app repackaging (O1:~\gou; O2:~\gou).

Using \name consists of two phases, i.e., the compile- and run-time phase as shown in \myfig~\ref{fig:seclib}.
Firstly in the compile-time phase, developers include \name into their app projects
and run our tool suite to help \name identify risky components that need protection.
Then limited amounts of stub codes are added into entry functions of risky components (usually two LOC per entry) so that \name can intercept incoming IPC flows.
Note that the whole procedure could be automatic with just a run of our tool suite.

In the run-time phase, \name automatically collects enforcement primitives and enforces policy checking without developers' involvement.
Considering that \name's checking is conducted only at entry points and only for risky components, it makes \name lightweight (O4:~\gou).
With \name, the incoming IPC flow no longer directly executes the component codes.
Instead, it has to first go through \name's checking that could generate three possible outputs: \texttt{deny}, \texttt{alert}, and \texttt{allow}.
Only in the \texttt{allow} case will the execution flow go back to the component code immediately.
In the case of \texttt{alert}, \name pops up an alert dialog for users to make a decision --- flow resumes if the users choose to allow the call.
If \name determines to \texttt{deny} to call, control will return to the calling environment immediately.

To make practical in-app policies to facilitate the access control, \name collects a number of enforcement primitives that previous efforts have not fully leveraged and takes all four types of Android components into consideration (O3:~\gou).
To achieve these objectives, we face the following major challenges in designing and implementing \name:

\begin{table*}[t!]
\centering
\caption{MAC policies in \name. Here are the six representative policies (P1 to P6) we have designed.}
\vspace{-3ex}
\scalebox{0.9}{
\begin{threeparttable}
\begin{tabu}{ |c | l | c | l | c|}

\hline
\rowfont{\bfseries}
\rowcolor{LightGray}
ID & Policy Name & \dag & Policy Representation & \ddag \tabularnewline
\hline
\hline

P1 & No By-default Exported Provider
& P
& \textbf{if} \blue{$\mathit{ID}_{a} \neq \mathit{ID}_{v}$} $\land$ \blue{$\neg \mathit{ExportedAttr}$}: \textbf{deny}
& \ling
\tabularnewline \hline

P2 & No Pre-claimed Custom Permission
& All
& \textbf{if} \blue{$\mathit{ID}_{a} \neq \mathit{ID}_{v}$} $\land$ \blue{$\exists\mathit{(PermAttr}_{v} \notin \mathit{SysPerms)}$} $\land$ \blue{$\mathit{PermAttr}_{v} = \mathit{PermAttr}_{a}$}: \textbf{deny}
& \ling
\tabularnewline \hline

\multirow{2}{*}{P3} & Alerting Implicitly Exported
& \multirow{2}{*}{A, S, R}
& \multirow{2}{*}{\textbf{if} \blue{
  $\mathit{ID}_{a} \neq \mathit{ID}_{v}$} $\land$ \blue{$\neg \mathit{ExportedAttr}$} $\land$ \blue{$\mathit{ActionAttr} \notin \mathit{SysActions}$
  }: \textbf{alert}}
& \multirow{2}{*}{\gou}
\tabularnewline
& Components with Custom Action & & & \tabularnewline
\hline

P4 & Alerting Explicitly Exported Provider
& P
& \textbf{if} \blue{$\mathit{ID}_{a} \neq \mathit{ID}_{v}$} $\land$ \blue{$\mathit{ExportedAttr} = \mathit{true}$}: \textbf{alert}
& \cha
\tabularnewline \hline

P5 & Checking System-only Broadcasts
& R
& \textbf{if} \blue{$\mathit{ID}_{a} \neq \mathit{ID}_{v}$} $\land$ \blue{$\exists \mathit{(ActionAttr} \in \mathit{SysActions)}$} $\land$ \blue{$\mathit{InputAction} \neq \mathit{ActionAttr}$}: \textbf{deny}
& \gou
\tabularnewline \hline

P6 & Filtering Sql Injection for Provider
& P
& \textbf{if} \blue{$\mathit{ID}_{a} \neq \mathit{ID}_{v}$} $\land$ \blue{$\exists \mathit{(AttackStr} \in \mathit{InputPara)}$}: \textbf{deny}
& \cha
\tabularnewline \hline

\end{tabu}
\begin{tablenotes}
\item[\dag] lists which components this policy is applicable to. All: all four components; A: \activity; S: \service; R: \receiver; and P: \provider.
\item[\ddag] indicates whether a policy has been covered by previous efforts. \gou = covers by~\cite{IntraComDroid12}; \cha = does not cover; \ling = partially covers by system updates. Note that \cite{IntraComDroid12} simply un-exports implicitly exported components in policy P3, which would cause incompatibility issues while ours will not. 
\end{tablenotes}
\end{threeparttable}
}
\label{tab:macpolicy}
\end{table*}

\begin{compactitem}
\item[C1]
\textit{It is challenging to design in-app policies in our context, an effort that had never been attempted before.}
Specifically, we need to cover all four types of components and mitigate attacks originated from both system weaknesses and common developer mistakes.
Our design is further restricted by the fact that the policies are enforced \textit{only} at entry points without additional run-time execution information.
Fortunately, as we are going to present in \mysec\ref{sec:policy}, it is \textit{feasible} to achieve a set of such in-app policies.

\item[C2]
\textit{Android fails to provide the caller identity information to most of callee components} except \provider and \service's Binder interface functions.
Although there is an API called \texttt{Binder.getCallingUID()} for this purpose, it only works in the Binder thread~\cite{Scippa14} in which most components' entry functions do not run.
Moreover, as an application-layer defense, \name cannot modify the Android OS to address this limitation, as opposed to Scippa~\cite{Scippa14}.
However, the caller identity is essential to the effectiveness of \name, which demands a solution for this challenge.

\item[C3]
\textit{It is nearly infeasible to pop up alert dialogs in the intercepted components}, because those entry functions lack appropriate user interface context and impose limited function return timing thresholds.
In \mysec\ref{sec:implementUI}, we will elaborate on this in more detail and how we overcome it by a dialog-like Activity transition technique.

\item[C4]
\textit{We do not have API support to collect certain component attributes}, though most of component attributes can be retrieved through existing APIs.
For example, Android simply reports \texttt{true} for all exported cases, no matter if it's explicitly or implicitly exported.
To obtain the raw \texttt{exported} value, \name performs run-time manifest analysis by itself, as shown in~\mysec\ref{sec:implementAttr} later.
\end{compactitem}



\subsection{In-app MAC Policy Design}
\label{sec:policy}

It is important to understand the policies \name is designed to enforce before we present other details.
Remember that our objective is to have mandatory access control (MAC) policies to stop common component hijacking issues that result from system flaws or developer mistakes.
Table~\ref{tab:macpolicy} lists six representative MAC policies (P1 to P6) we have designed.
From a high-level view, policies P1 and P2 patch the system weaknesses, P3 to P5 mitigate common developer mistakes, and P6 filters a common attack.
Note that we do not claim that they cover all hijacking issues; instead, our purpose is to show \textit{how to design} in-app \name policies for major categories of attacks and for different components.
Our policies thus serve as templates or baselines for more enhanced or customized policies.
Researchers can easily make other policies as we will demonstrate below.

Before going to the policy details, we first introduce the symbols used in these policies.
We denote each attribute as $\mathit{XxxAttr}$, such as $\mathit{ExportedAttr}$ for the exported attribute, $\mathit{PermAttr}$ for the permission attribute, and $\mathit{ActionAttr}$ for the \intent actions registered by components.
Additionally, there are a number of system-defined permissions and actions, which we denote as $\mathit{SysPerms}$~\cite{SysPerms} and $\mathit{SysActions}$~\cite{SysActions}, respectively.
If a permission can be used only by system or an action can be sent only by system, we further denote them as $\mathit{SysOnlyPerm}$ and $\mathit{SysOnlyAction}$, respectively.
In addition, we denote each input data as $\mathit{InputXxx}$.
For example, the incoming \intent action is represented as $\mathit{InputAction}$ and the data \texttt{Uri} is denoted as $\mathit{InputUri}$.

\textbf{Trusting intra-app IPC by default.}
A common point among all six policies is that we consider the IPC calls initiated from the same app/developer trusted.
That is, only an external IPC call from a third-party app will be checked, i.e., $\mathit{ID}_{a} \neq \mathit{ID}_{v}$.
This by-default rule is important in two aspects.
First, it effectively minimizes the usability issues for normal user operations, because only the \textit{external} IPC for certain exported components (i.e., risky components, as we will present in \mysec\ref{sec:other}) will trigger the alerts.
Second, it strengthens \name's access control capabilities because another app from the same developer now can be trusted through the app identity and its developer certificate checking.
In contrast, solutions such as IntraComDroid~\cite{IntraComDroid12} and Android Lint simply stop all incoming IPC calls, including those from the same developer, by un-exporting components.

However, if the victim app contains a ``middle'' component that allows an experienced adversary to detour external IPC requests first to that middle component, our policy for trusting intra-app IPC could be bypassed. 
Although it's unusual for the victim app to have this middle-component problem, we still can mitigate it by adding a flag into each IPC relay so that \name can infer the origin of an IPC call chain.
For example, we can add one line of code, \texttt{intent.putExtra(`sclibflag', `outside')}, into the Intents from outside.
We leave how to robustly address this issue to our future work.
In this paper, \name takes the default setting that trusts all intra-app IPC.

\textbf{Fixing system weaknesses with P1 and P2.}
We now show how to design \name policies to mitigate system flaws.
To this end, we design policies P1 and P2 to fix two major weaknesses in Android.
The first is that Android prior to 4.2 by default exports all \provider components even if they do not claim the \texttt{exported} attribute.
This over-ambitious exposure rule led to thousands of vulnerable \provider components~\cite{ContentScope13}.
Although Google later disabled it, there are still many by-default exported \provider components in apps compiled with old SDK according to recent studies~\cite{TargetFragmentation16, DSDK17}.
To make the new rule available to all apps and all phones (including those under 4.2), we design policy P1 to mimic the current system rule at the app layer.
Specifically, \name directly denies an external IPC request to the callee \provider that does not claim the \texttt{exported} attribute ($\neg \mathit{ExportedAttr}$).

P2, on the other hand, fixes a more complicated and less-known system flaw~\cite{CustomPermission} where an attack app installed earlier can pre-claim a custom permission in the victim app with the purpose of downgrading its protection level, e.g., from \texttt{signature} to \texttt{normal}.
Consequently, the attack app can hijack a ``private'' component that was originally protected with \texttt{signature}-level permissions.
Based on this root cause, policy P2 first determines whether there is a custom permission defined in the callee component, i.e., $\exists(\mathit{PermAttr}_{v} \notin \mathit{SysPerms})$.
If it exists, we further check whether or not it has been pre-claimed by the caller app, i.e., $\mathit{PermAttr}_{v}$ $ = \mathit{PermAttr}_{a}$.
In practice, we can simplify policy P2 for the \texttt{signature}-level custom permissions by leveraging the fact that an external IPC can come only when the \texttt{signature} permission has been downgraded.

\textbf{Preventing developer mistakes with P3 to P5.}
In this part, we show that how \name prevents three common developer mistakes using policy P3 to P5.
We first discuss policy P3 and P4 to take care of developers who mistakenly export their components or simply did not realize the threats from exported components.
Specifically, for policy P3, our insight is that if developers register custom \intent actions for their implicitly exported components, very likely they do not intend to export those components. 
While policy P4 is based on the measurement results in~\cite{ContentScope13, TargetFragmentation16} that many explicitly exported \provider components can also leak sensitive data.
To prevent these two types of mistakes, policy P4 checks $\mathit{ExportedAttr}$ and policy P3 further checks the custom actions ($\mathit{ActionAttr} \notin \mathit{SysActions}$).
To reduce false positives, we choose the \texttt{alert} for policy P3 and P4.
Moreover, since custom actions and \provider components are much less-called by inter-app IPC, we expect that our \texttt{alert} policies would not disrupt user experience.

Then we have policy P5 to mitigate a developer mistake that appears at the code level (instead of manifest).
That is, a \receiver component that registers system-only broadcasts is still hijack-able if it does not check the incoming \intent action explicitly in the code~\cite{ECVDetector14}.
Our measurement of ten high-profile open source apps in \mysec\ref{sec:evaluate} shows that a couple of them made this mistake.
To defend, policy P5 automatically checks the input action (on behalf of callee component) against the system-only action claimed in manifest, i.e., $\exists (\mathit{ActionAttr} \in \mathit{SysAction})$ $\land$ $\mathit{InputAction} \neq \mathit{ActionAttr}$.

We can devise more policies to prevent other mistakes.
For example, we can also mitigate the denial-of-service hijacking due to missed null checks on IPC input.
To stop such hijacking, we record the app crash times corresponding to each caller app, and if it has exceeded the threshold value (e.g., three times), \name then denies further requests.
We leave implementing such a policy as our future work.

\textbf{Stopping a common attack with P6.}
Finally, we propose policy P6 as a prominent example to show \name's capability of stopping common attacks.
Specifically, policy P6 aims to filter SQL injection for \provider.
As demonstrated in~\cite{ContentScope13}, an attack app can hijack a \provider component to inject malicious SQL statements.
For example, the adversary sets the \texttt{projection} parameter of the \texttt{query} function as a special phase ``\texttt{* from private\_table;}''.
As these special inputs are different from normal queries, we use keyword-based filtering (such as the expression like ``\texttt{xxx from yyy;}'') to stop them.
Similarly, we can stop the directory traversal attack~\cite{ContentScope13} in \texttt{openFile} entry of \provider by leveraging some file path patterns.
Furthermore, we can devise \texttt{alert} policies to protect permission-protected components to stop an adversary that claims corresponding \texttt{dangerous} permissions, as we will conduct case studies in \mysec\ref{sec:security}.

\subsection{Compile-time Designs}
\label{sec:other}

\begin{table*}[t!]
\centering
\caption{Component entry functions that need to be intercepted, characterized by component types and IPC caller APIs.}
\vspace{-3ex}
\scalebox{0.9}{
\begin{threeparttable}
\begin{tabu}{ |c | c | l|}

\hline
\rowfont{\bfseries}
\rowcolor{LightGray}
Component types & IPC caller APIs\tnote{\dag} & Interested entry functions\tnote{\dag} of IPC callee components \tabularnewline
\hline
\hline

\multirow{2}{*}{\activity} & \texttt{Context.startActivity()} & \multirow{2}{*}{\texttt{onCreate(Bundle)}; \texttt{onStart()}; \texttt{onNewIntent(Intent)}}\tabularnewline
& \texttt{Activity.startActivityForResult()} & \tabularnewline
\hline

\multirow{2}{*}{\service} & \texttt{Context.startService()} & \texttt{onCreate()}; \texttt{onStartCommand()}\tnote{\ddag} ; \texttt{onHandleIntent(Intent)} \tabularnewline
& \texttt{Context.bindService()} & \texttt{onCreate()}; \texttt{onBind(Intent)}; \texttt{onRebind(Intent)} \tabularnewline
\hline

\multirow{2}{*}{\receiver} & \texttt{Context.sendBroadcast()} & \multirow{2}{*}{\texttt{onReceive(Context, Intent)}} \tabularnewline
& \texttt{Context.sendOrderedBroadcast()} & \tabularnewline
\hline

\multirow{6}{*}{\provider} & \texttt{ContentResolver.query()}  & \texttt{query(Uri, String[], String, String[], String)} \tabularnewline
& \texttt{ContentResolver.insert()}  & \texttt{insert(Uri, ContentValues)} \tabularnewline
& \texttt{ContentResolver.bulkInsert()}  & \texttt{bulkInsert(Uri, ContentValues[])} \tabularnewline
& \texttt{ContentResolver.update()} & \texttt{update(Uri, ContentValues, String, String[])} \tabularnewline
& \texttt{ContentResolver.delete()}  & \texttt{delete(Uri, String, String[])} \tabularnewline
& \texttt{ContentResolver.openFileDescriptor()}  & \texttt{openFile(Uri, String)} \tabularnewline
\hline

\end{tabu}
\begin{tablenotes}
\item[\dag] For simplicity, we skip the parameters of caller APIs and the class names of entry functions.
\item[\ddag] An old-SDK version of \texttt{onStartCommand(Intent, int, int)} is \texttt{onStart(Intent, int)}, which should be also covered.
\end{tablenotes}
\end{threeparttable}
}
\label{tab:entryfuncs}
\end{table*}

At compile-time, \name mainly identifies the risky components and their entry functions in order to add stub code for policy checking and enforcement.

\noindent\textbf{Identifying risky components.}
In order to minimize the overhead introduced by \name, we only add our checking to {\em risky} components that are defined as
\begin{compactitem}
\item An \activity component is risky if it is exported and registers custom \intent actions. 

\item A \service component is risky if it is (i) explicitly exported with custom \intent actions; or (ii) implicitly exported.

\item A \receiver component is risky if it is (i) explicitly exported with custom/system-only \intent actions; or (ii) implicitly exported.

\item A \provider component is risky if it does not claim the \texttt{exported} attribute being \texttt{false}.
\end{compactitem}

\noindent\textbf{Locating entry functions.}
After risky components are identified, \name makes MAC policy decisions at the beginning of each call, and therefore needs to identify entry functions of the IPC calls.
Table~\ref{tab:entryfuncs} characterizes the component entry
functions that need to be intercepted. They are organized by
different component types and IPC call APIs. Due to space
limitation, we briefly explain these entry functions in the
reverse order of component types:
\begin{compactitem}
\item[\textit{\provider}:]
Identifying entry functions for \provider is straightforward, because there is a one-to-one mapping between each caller API and entry function.

\item[\textit{\receiver}:]
There is only one entry for \receiver, namely the \texttt{onReceive} callback function.

\item[\textit{\service}:]
There are two ways to call a \service component, either by starting it via the \texttt{startService} API or by binding it via the \texttt{bindService} API.
The \texttt{onStartCom-} \texttt{mand} entry (in the first case) operates in a way similar to the \texttt{onReceive} function, whereas the \texttt{onBind} or \texttt{onRebind} entry (in the second case) only returns a \binder object and has no further sequential execution.
Indeed, with the retrieved \binder object, an adversary can invoke any IPC interface functions~\cite{BoundService, AIDL} that are pre-defined by the \binder object.
These custom interface functions thus become additional \service entries.
Additionally, \texttt{onHandleIntent} is the entry of the subclass \texttt{IntentService}.

\item[\textit{\activity}:]
Similar to \service, the entry functions of \activity also include \texttt{onCreate} and \texttt{onStart}.
Moreover, \texttt{onNewIntent} could be an additional entry.
\end{compactitem}

\noindent\textbf{Adding stub code.}
To route incoming IPC flows to \name for access control, \name adds two lines of stub code into entry functions. 
Listing~\ref{lst:entryactivity} demonstrates how we instrument \activity's \texttt{onCreate} function.
Instrumenting other components' entry functions is generally similar or could be even easier.
For example, we instrument \receiver's \texttt{onReceive} via \texttt{SecReceiver.receive(context, intent)}, and handle \provider's \texttt{query} by intercepting all its parameters plus a \texttt{this} variable for \texttt{Context}.

\begin{lstlisting}[
float=h!,
style=customJava,
basicstyle=\ttfamily\scriptsize,
label={lst:entryactivity},
caption={Instrumenting \texttt{Activity}'s \texttt{onCreate}, with the additional \texttt{getIntent()} and \texttt{finish()} API calls.}
]
protected void onCreate(Bundle savedState) {
    super.onCreate(savedInstanceState);
+   if (!SecActivity.onCreate(this, getIntent()))
+   {   finish();   return;     }
    // Original code
}
\end{lstlisting}

A special case is how we handle bound services that own Binder interface functions.
As mentioned above, we shall treat those interfaces as individual entries.
Therefore, for a bound \service such as the RemoteService~\cite{AIDL} (see its code at \url{http://tinyurl.com/getpidentry}), we instrument not only its \texttt{onBind} as usual, but also its \texttt{getPid} interface in a way similar to how we instrument \texttt{Provider}'s entries.
Specifically, we record both the interface name (``getPid'') and all its parameters (\texttt{null} in this case).

\subsection{Recovering Caller Identity via the Binder Side Channel}
\label{sec:implementCaller}

\begin{figure}[t!]
\begin{adjustbox}{center}
\includegraphics[width=0.45\textwidth]{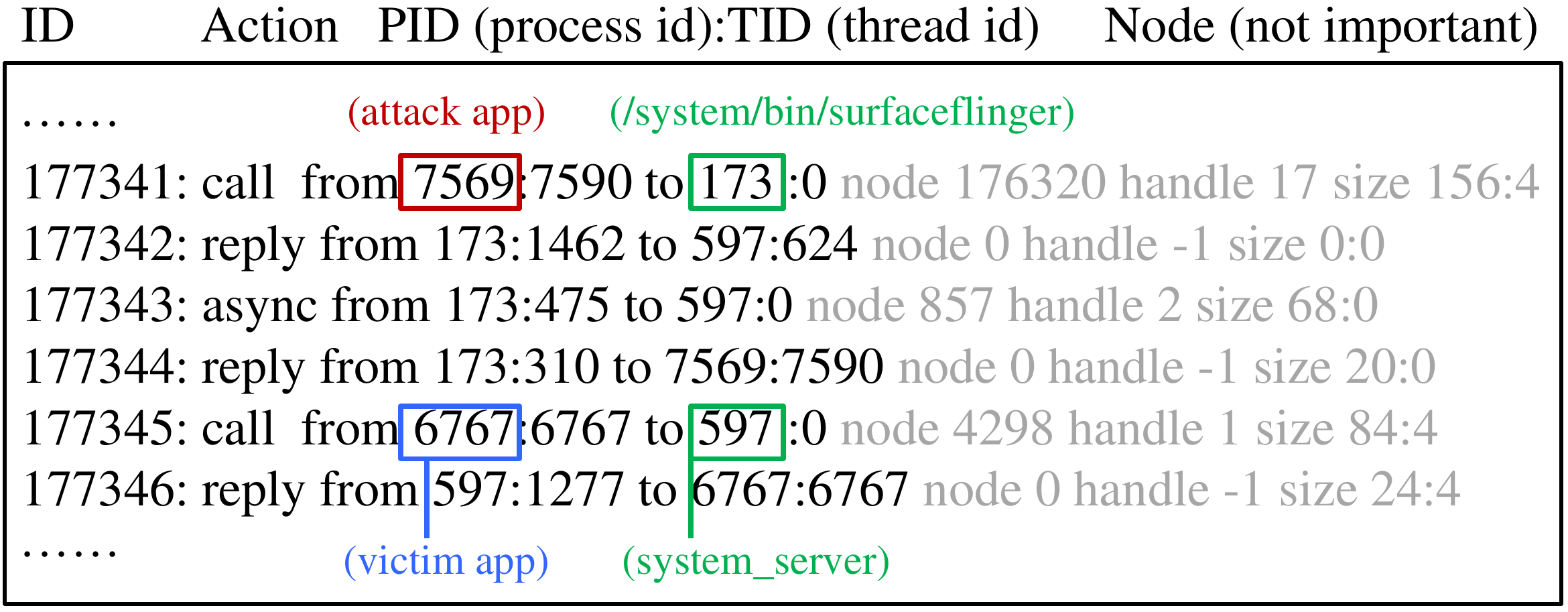}
\end{adjustbox}
\vspace{-5ex}
\caption{An example of the Binder transaction\_log.}
\label{fig:binderlog}
\vspace{-4ex}
\end{figure}

Having discussed the policies that \name is designed to enforce, we now turn to some implementation details to show how \name managed to overcome the design challenges.  In this specific subsection, we show how \name recovers the caller identify (C2) via the Binder side channel at the path of \texttt{/sys/kernel/debug/binder/transac} \texttt{tion\_log}.
More specifically, for each risky IPC call intercepted, we retrieve the recent Binder logs from this side channel and analyze them to recover the corresponding caller app identity.
\myfig~\ref{fig:binderlog} shows a \texttt{transaction\_log} example when an attack app exploits an \activity component in the victim app.
Each Binder log starts with a unique transaction ID followed by the Binder action and the process/thread IDs of the caller and callee processes.
The last part, node information, is not important --- so we skip here.
Note that in the kernel-layer Binder driver, app processes do not directly interact with each other.
Instead, the high-level IPC always involves a number of interactions between apps and system processes (see~\cite{Scippa14} for more details).
For example, the attack app (PID: 7569) and the victim app (PID: 6767) here leverage the \texttt{surfaceflinger} and \texttt{system\_server} processes to delegate their communication.

Our extensive tests of Binder logs in different components show that they all follow the same pattern, based on which we propose a simple yet effective algorithm to recover caller identities.
We still use \myfig~\ref{fig:binderlog} to illustrate this algorithm.
The first step is to locate the Binder log that ``calls from'' the callee PID for the first time, i.e., the transaction 177345.
Then we trace back to identify the first app process, i.e., PID 7569, which is the caller app we are looking for.
Since there is no fixed PID pattern for non-system processes, we further extract the corresponding UID and package name for analysis.
More specifically, if the UID is smaller than 10000 or if the package name is a system binary, it must be a system process.

Since there is a timing window to retrieve the recent Binder logs, \name performs the Binder analysis before other modules.
To further decrease the delay, we focus on extracting and saving the logs first, and postpone the actual analysis.
Our tests show that in this way, we can reliably retrieve the required Binder logs.
Moreover, we found that accessing the Binder transaction log is allowed even in some smartphones with SEAndroid.
We tested more than ten Android device models and found that the majority of them allow the access in the SEAndroid enforcing mode, including Samsung Galaxy S6 Edge+, Nexus 4/5/5X/6P, and several Huawei/Samsung/XiaoMi phones.

\subsection{Popping Up Alerts via the Dialog-like Activity Transition}
\label{sec:implementUI}

To enforce the \texttt{alert} policies, \name needs to pop up an alert dialog for users to choose ``allow'' or ``deny''.
However, this is a challenging task due to the following reasons:
\begin{compactitem}
\item Background components such as \provider and \service do not have an appropriate UI \texttt{Context} to display alert dialogs.
Even for \activity, it cannot pop up dialogs when \texttt{onCreate()} is still being intercepted, (i.e., not return yet).

\item Some components' entry functions (e.g., \activity's \texttt{onCreate()} and \receiver's \texttt{onReceive()}) need to return in a short time.
Therefore, we cannot hold on the execution of these functions and wait for users' decisions.
\end{compactitem}

To address these issues, we opt for a different strategy instead of directly displaying an alert dialog.
The basic idea is to launch a dialog-like \activity from the intercepted component via the \texttt{startActivity()} API.
For entry functions that are sensitive to execution time, \name immediately returns the execution flow to them by assuming users choosing ``deny''.
If users select ``allow'' later, \name re-sends the same Intent\footnote{\small \provider's entry functions are not sensitive to execution time, so we focus on Intent-based communications here.} content on behalf of the original caller app.
Since the callee app has no way to distinguish the caller identity, the original execution flow can resume.
While for other time-insensitive entry functions, \name can pause their component execution and wait for users to make a decision on the alert dialog.


To implement this dialog-like Activity transition approach, we identify and handle the following technical issues:
\begin{compactitem}
\item To enable background components to be able to launch the alert \activity, we set a special \texttt{Intent} flag called \texttt{FLAG\_ACTI} \texttt{VITY\_NEW\_TASK}.

\item After users click ``allow'' or ``deny'', we automatically return to the original UI state.
This is important for maintaining a good user experience.
To do so, we set one more \intent flag called \texttt{FLAG\_ACTIVITY\_MULTIPLE\_TASK} so that our alert \activity can naturally go back to the previous UI state after calling \texttt{Activity.finish()}.

\item To avoid alert dialogs appearing in the history of recent \activity components, we set the third \intent flag called \texttt{FLAG\_ACTIVITY\_EXCLUDE\_FROM\_RECENTS}.

\item To pause the component execution and wait for user decisions, \name initializes a lock object and sets this object into the ``wait'' status after invoking \texttt{startActivity()}. Once users click the buttons, the alert \activity will notify the ``waited'' lock object to be released. Then, the lock object will stop waiting and the paused component will resume its execution. To avoid unnecessarily long waiting, we set a timeout value of 30 seconds in our prototype.
\end{compactitem}

%


\subsection{Extracting Component Attributes by Run-time Manifest Analysis}
\label{sec:implementAttr}

To overcome challenge C4 that we do not have API support to collect certain component attributes, \name performs Android manifest analysis by itself.
In particular, we choose the run-time analysis instead of compile-time analysis because it does not bother developers and neither needs the additional file storage.
Also, it is immune to app updates and can handle new components well.

The basic procedure is to \textit{dynamically} retrieve and parse \texttt{Android} \texttt{Manifest.xml} of the callee app.
Specifically, for the callee component, we extract its raw \texttt{exported} status, the registered Intent actions, and the associated permissions.
We then correlate the app permission entries to obtain their protection levels and determine whether an associated permission is defined by the system or the callee.
We also build a list of system-defined and system-only Intent actions based on the Stowaway result~\cite{Stowaway11} and Android source code so that we can determine whether a given component listens to system Intent actions or not.






\section{Evaluation}
\label{sec:evaluate}

In this section, we evaluate \name in three aspects.
Firstly in \mysec\ref{sec:tenapps}, we measure the component statistics of ten high-profile open source apps to find out how many risky components could benefit from \name and how much code footprint \name introduces.
Then in \mysec\ref{sec:security}, we assess the security effectiveness of \name against attacks in different components.
Finally in \mysec\ref{sec:performance}, we measure the performance overhead of \name under different scenarios.

\subsection{Applying \name}
\label{sec:tenapps}

\begin{table*}[t!]
\centering
\caption{Detailed statistics of ten high-profile open source applications.}
\vspace{-3ex}
\label{tab:dataset}
\scalebox{0.85}{
\begin{threeparttable}
\begin{tabular}{|c|c|c|c|c|c|c|c|c|c|c|}
\hline
Application           & Installs     & Version  & Start date  & \begin{tabular}[c]{@{}c@{}}\# of\\  Activity\dag\end{tabular} & \begin{tabular}[c]{@{}c@{}}\# of\\  Service\dag\end{tabular} & \begin{tabular}[c]{@{}c@{}}\# of\\  Receiver\dag\end{tabular} & \begin{tabular}[c]{@{}c@{}}\# of \\ Provider\dag\end{tabular} & \begin{tabular}[c]{@{}c@{}}\# of \\ custom \\ permissions\end{tabular} & \begin{tabular}[c]{@{}c@{}}\# of \\ components\\ with custom \\ permissions\ddag\end{tabular} & \begin{tabular}[c]{@{}c@{}}\# of\\ risky\\ components\ddag\end{tabular} \\ \hline

Telegram              & 100M - 500M  & 3.13.1   & Oct, 2013   & (1,2)/6                                                   & (4,1)/9                                                  & (0,4)/10                                                  & (0,0)/1                                                              & 0                                                                            & 0,1,0,0                                                                                                  & 0,1,4,0                                                            \\ \hline

Zxing Barcode & 100M - 500M  & 4.7.7    & Oct, 2011  & (0,4)/9                                                   & (0,0)/0                                                  & (0,0)/0                                                   & (0,0)/0                                                              & 0                                                                           & 0,0,0,0                                                                                                  & 4,0,0,0                                                            \\ \hline

Terminal Emulator     & 10M - 50M    & 1.0.70   & before 2010 & (1,5)/8                                                   & (0,1)/1                                                  & (0,0)/0                                                   & (0,0)/0                                                              & 3                                                                           & 1,0,0,0                                                                                                  & 3,1,0,0                                                            \\ \hline

K-9 Mail              & 5M - 10M     & 5.010    & before 2010 & (0,7)/27                                                  & (0,0)/7                                                  & (0,4)/5                                                   & (2,0)/4                                                              & 4                                                                           & 0,1,1,2                                                                                                  & 1,0,4,2                                                            \\ \hline

WordPress             & 5M - 10M     & 5.9      & Feb, 2012   & (0,10)/53                                                 & (0,0)/15                                                 & (1,2)/5                                                   & (0,0)/1                                                              & 1(S)                                                                           & 0,0,0,0                                                                                                  & 1,0,2,0                                                            \\ \hline

Signal Messenger & 1M - 5M      & 3.20.4   & May, 2010   & (1,4)/37                                                  & (3,0)/8                                                  & (5,5)/12                                                  & (0,0)/2                                                              & 2(S)                                                                     & 0,0,0,0                                                                                                  & 0,0,5,0 \\ \hline

Wire                  & 1M - 5M      & 2.19.289 & Dec, 2014   & (0,1)/5                                                   & (2,0)/2                                                  & (0,1)/1                                                   & (0,0)/0                                                              & 0                                                                           & 0,0,0,0                                                                                                  & 0,0,1,0                                                            \\ \hline

Bitcoin Wallet        & 1M - 5M      & 5.0      & Mar, 2011   & (1,2)/11                                                  & (0,0)/4                                                  & (0,2)/2                                                   & (0,0)/3                                                              & 0                                                                           & 0,0,0,0                                                                                                  & 0,0,2,0                                                            \\ \hline
ChatSecure            & 0.5M - 1M    & 14.2.3   & Feb, 2011   & (0,2)/25                                                  & (0,0)/5                                                  & (2,0)/2                                                   & (0,0)/1                                                              & 3                                                                           & 0,1,0,1                                                                                                  & 1,0,0,0                                                            \\ \hline
Zirco Browser            & 50K - 100K      & 0.4.4    & Jun, 2010 & (0,1)/17                                                  & (0,0)/0                                                  & (0,1)/1                                                   & (0,2)/2                                                              & 0                                                                           & 0,0,0,0                                                                                                  & 0,0,1,2                                                            \\ \hline
\end{tabular}
\begin{tablenotes}
\item [\dag] \textit{(x,y)/z} in column 5, 6, 7, and 8 means the app has a total of \textit{z} components of the type, among which \textit{x} are explicitly exported and \textit{y} are implicitly exported. The rest are non-exported.
\item [\ddag] The data format in column 10 and 11, \textit{a,b,c,d}, follows the order of \activity, \service, \receiver, and \provider. 
\item [(S)] The number with the \textit{(S)} mark in column 9 indicates \texttt{signature}-level permissions. The other permissions in this column are at the \texttt{dangerous} level.
\end{tablenotes}
\end{threeparttable}
}
\end{table*}

We first get an idea about the extent to which typical Android apps export their components to others, and the corresponding code footprint when we apply \name to protect these components.
To this end, we collect the latest source code of ten high-profile open source apps from their GitHub sites at the time of our research (November 2016).
Note that since the intended users of \name are developers who own codes, \name is not only applicable to open source apps but also regular apps on Google Play.

We summarize the statistics of ten tested apps in Table~\ref{tab:dataset}.
The majority of them are popular, with millions of downloads according to Google Play, and the development of most apps have been lasting for over five years.
Furthermore, several of them claim to be highly secure, including the very popular Telegram, K-9 Mail, and Signal Private Messenger.
Additionally, although Zirco Browser only has around 100K installs, it is used by several other browser apps (e.g., \texttt{org.easyweb.browser} with 1M--5M installs) as a baseline.

We then run \name's scripts to automatically measure the exported and risky components as well as their custom permissions.
The measurement results are also shown in Table~\ref{tab:dataset}, classified by different component types.
We find that every tested app exports some of its components, and 67.3\% of the exported components are implicitly exported.
Moreover, a large portion of the exported components are potentially risky, which yield a total of 35 risky components that require \name's protection.
Furthermore, we find that five apps define their own custom permissions for their components, which makes them potential victims to the permission pre-occupy attack (see policy P2 in \mysec\ref{sec:policy}).

We further measure the additional stub code introduced by \name.
Specifically, we calculate the number of additional lines of code based on the type of risky component and the number of entry functions of that type.
The results in Table~\ref{tab:stub} show that the code footprint introduced by \name is negligible at less than 0.3\% in all cases. 
Note that since \name is implemented in Java and will be instrumented in Java environments, we only compare our code footprint based on the number of lines of Java code in each apps, though some tested apps also contain many C/C++ codes.
Additionally, the jar file of \name itself is also very small --- only around 30KB before compression.

\begin{table}[t!]
\centering
\caption{Size of stub code to protect risky components.}
\vspace{-3ex}
\label{tab:stub}
\scalebox{0.9}{
\begin{tabular}{|c|c|c|c|}
\hline
Application           & \begin{tabular}[c]{@{}c@{}}Lines of \\ Java codes\end{tabular} & \begin{tabular}[c]{@{}c@{}}Lines of \\ stub codes\end{tabular} & \begin{tabular}[c]{@{}c@{}}Extra code \\ percentage\end{tabular} \\ \hline
Telegram              & 222,074                                                       & 32                                                            & 0.014\%                                                          \\ \hline
Zxing Barcode  & 43,221                                                        & 24                                                            & 0.056\%                                                          \\ \hline
Terminal Emulator     & 11,507                                                        & 30                                                            & 0.261\%                                                          \\ \hline
K-9 Mail              & 51,416                                                        & 62                                                            & 0.121\%                                                          \\ \hline
WordPress             & 81,076                                                        & 22                                                            & 0.027\%                                                          \\ \hline
Signal Messenger                & 63,137                                                        & 34                                                            & 0.054\%                                                          \\ \hline
Wire                  & 52,808                                                        & 2                                                             & 0.004\%                                                          \\ \hline
Bitcoin Wallet        & 18,695                                                        & 40                                                            & 0.214\%                                                          \\ \hline
ChatSecure            & 36,911                                                        & 18                                                            & 0.049\%                                                          \\ \hline
Zirco Browser         & 9,638                                                          & 26                                                             & 0.270\%                                                          \\ \hline
\end{tabular}
}
\vspace{-2ex}
\end{table}

\subsection{Security Evaluation}
\label{sec:security}


To perform security evaluation, we identify eight vulnerable or risky components from the aforementioned ten apps.
As shown in Table~\ref{tab:CaseStudy}, these cases cover all four types of Android components and all six policies we designed.
We first implement two attack apps for the Intent-based and \provider-based attacks, and verify that such attacks work on the (original) vulnerable or risky components.
Our tests show that after including \name, the security-enhanced components are no longer vulnerable to those attacks.

In the rest of this subsection, we present our detailed analysis of the eight case studies to demonstrate \name's unique values in mitigating developer mistakes and system weaknesses when compared to developers' own patches and Android platform updates.

\textbf{Case 1: Fixing vulnerable components without losing compatibility.}
The first case, Terminal Emulator (\texttt{jackpal.androidterm}), is a good example to illustrate that the developers' own patches sometimes could cause incompatibility issues that \name can avoid.
Terminal Emulator contained a vulnerable component called \texttt{Remote} \texttt{Interface} in its version 1.0.63.
The component is implicitly exported and can be triggered by a crafted Intent to execute arbitrary commands without any user interaction.
To fix this vulnerability, the developers removed the programmatic command execution functionality in \texttt{RemoteInterface}~\cite{TermFix, TermIssue}.
However, there were other apps\footnote{\small To name a few, see code snippets in \url{https://goo.gl/HK3HgJ}, \url{https://goo.gl/0t78J8}, \url{https://goo.gl/xPjlv3}, and \url{https://goo.gl/xOs5zN}.} that continue to utilize this programmatic interface and the patch thus caused an incompatibility issue\footnote{\small A bug report was actually issued after the patch, but the developers of Terminal closed all links after the project was finished.} on those apps.
Additionally, simply un-exporting the component as proposed by IntraComDroid~\cite{IntraComDroid12} would cause the same incompatibility issue.

In contrast, \name fixes this vulnerability in a more elegant way that results in no incompatibility issue and no additional developer effort.
Specifically, since \texttt{RemoteInterface} registers an Intent filter to take a custom Intent action, it satisfies our policy P3 (see Table~\ref{tab:macpolicy}).
As a result, \name pops up an alert dialog when an external app tries to trigger the programmatic command execution in \texttt{RemoteInterface}.
In this way, \name notifies users on potential attacks while keeping the app compatible with other legacy apps (that call \texttt{RemoteInterface}).
\name also saves the developers' effort in making the patches --- Terminal's developers performed around 200 lines of code changes to construct the patch~\cite{TermFix}.

\begin{table}[t!]
\centering
\caption{Security case studies: Using \name to protect vulnerable/risky components.}
\vspace{-3ex}
\label{tab:CaseStudy}
\scalebox{0.9}{
\begin{threeparttable}
\begin{tabu}{ |c | c | c | c|}

\hline
\rowfont{\bfseries}
\rowcolor{LightGray}
ID & Target Component (\dag) & App & Policy \tabularnewline
\hline
\hline

1
& \texttt{RemoteInterface} (A; I)
& Term Emulator
& P3 (alert)
\tabularnewline \hline

2
& \texttt{MessageProvider} (P; E)
& K-9 Mail
& P4 (alert)
\tabularnewline \hline

3
& \texttt{RemoteControlReceiver} (R; I)
& K-9 Mail
& P3 (alert)
\tabularnewline \hline

4
& \texttt{TermService} (S; I)
& Term Emulator
& P3 (alert)
\tabularnewline \hline

5
& \texttt{ZircoBookmarksProvider} (P; I)
& Zirco Browser
& P1 (deny)
\tabularnewline \hline

6
& \texttt{New/Clear KeyReceiver} (R; I)
& Signal
& P2 (deny)
\tabularnewline \hline

7
& \texttt{AppStartReceiver} (R; I)
& Telegram
& P5 (deny)
\tabularnewline \hline

8
& \texttt{WeaveContentProvider} (P; I)
& Zirco Browser
& P6 (deny)
\tabularnewline \hline

\end{tabu}
\begin{tablenotes}
\item [\dag] means \texttt{Type;} \texttt{Export}, i.e., the component type (four types of components) and the export status (implicitly or explicitly exported).
\end{tablenotes}
\end{threeparttable}
}
\vspace{-3ex}
\end{table}

\textbf{Case 2 \& 3 \& 4: Enforcing security beyond Android's existing security mechanisms.}
In this part, we first present how \name enhances protection of two risky components in K-9 Mail (\texttt{com.fsck.k9}) --- \texttt{MessageProvider} as in case 2 and \texttt{RemoteControl} \texttt{Receiver} as in case 3.
Both components are exported and have self-defined \texttt{dangerous}-level permissions.
The rationale behind this design is that K-9 Mail has a number of extension apps~\cite{WorkK9Email} which need to access these two components.
To share components to other apps with different signatures, the most secure way Android currently provides is to define a \texttt{dangerous}-level permission, as what K-9 Mail did.
However, this is too coarse-grained and cannot prevent a malicious app from claiming the corresponding permissions to steal users' emails via \texttt{MessageProvider}.
Indeed, according to a comprehensive survey~\cite{PermissionSurvey12}, users generally skip the permission inspection during app installation or simply cannot understand the permission meanings, which makes the attacks here realistic.

With \name, K-9 Mail now can achieve a more fine-grained access control by enabling users allow/deny a \textit{particular} external app on the alert dialog.
K-9 Mail would not have been capable of achieving such fine-grained security because:
(i) Intent-based components such as \texttt{RemoteControlReceiver} have no existing method of obtaining caller identity, an important primitive Android currently fails to provide;
and (ii) even though \texttt{MessageProvider} has an API to extract the caller identity, it cannot pop up alert dialogs.

Further, \texttt{TermService} in case 4 demonstrates a clearer example where developers actually demand the capability of differentiating different caller app identities.
According to its code at \url{http://tinyurl.com/termservice}, we see that the developers want to determine whether an external app or its own \activity makes the incoming IPC.
However, \texttt{TermService} tries to achieve this by checking whether the incoming Intent contains a custom action that is claimed in the \texttt{<intent-filter>}.
Developers believe that an external app would use that custom action to launch IPC, but actually an attack app can explicitly call \texttt{TermService} without setting that action.
Consequently, \texttt{TermService}'s action-based checking can be bypassed.
With \name, we can prevent such attacks and provide developers a solid mechanism to differentiate external IPC calls.


\textbf{Case 5 \& 6: Fixing system weaknesses with a broader platform and app coverage.}
Next, we introduce two cases to illustrate that \name can fix system weaknesses with a broader platform and app coverage than Android's system updates.
In case 5, Zirco Browser (\texttt{org.zirco})'s \texttt{ZircoBookmarksProvider} is by default exported by Android system, causing the leakage of users' bookmarks.
Although Android changed this by-default policy since 4.2, the new exposure policy is not applicable to apps with a target SDK version below 4.2.
In contrast, \name leverages the policy P1 to protect all implicitly exported \provider components even when they run on legacy phones or are compiled with target SDKs of older versions.

As another example, Signal Private Messenger (\path{org.thoughtcrime.securesms}) contains two dynamically registered \receiver components, \texttt{NewKeyReceiver} and \texttt{ClearKeyReceiver}, which are protected with a custom \texttt{signature}-level permission called \path{ACCESS_SECRETS}.
As explained in \mysec\ref{sec:policy}, they are subject to the permission pre-occupy attack.
Android fixes this weakness only after 5.0, whereas \name can eliminate its impact even on Android versions prior to 5.0.

\textbf{Case 7 \& 8: Fixing common developer mistakes and stopping common attack patterns.}
We now present case 7 and 8 to illustrate how \name helps fix a common developer mistake and stop a common attack pattern, respectively.
In case 7, Telegram (\texttt{org.telegram.messenger}) defines an \texttt{AppStartReceiver} component to listen to the \texttt{BOOT\_COMPLETED} broadcast, but the developers forgot to check this system-only action in its code (see \url{http://tinyurl.com/startreceiver}), making it possible that the component execution be triggered by any app.
With \name, developers no longer need to worry about such checking because \name automatically performs the checking based on policy P5.
We further mimic a SQL injection attack on \texttt{WeaveContentProvider} in case 8, which can be defended by our policy P6, as shown in \url{http://tinyurl.com/sqlweave}.

%
%

%
%


\subsection{Performance Evaluation}
\label{sec:performance}

We now evaluate the performance overhead of \name under different scenarios.
Since we have already shown that the code footprint \name introduces is very small (less than 0.3\%, see \mysec\ref{sec:tenapps}), here we focus our evaluation on the additional execution time \name introduces.

\textbf{Evaluation methodology.}
As the additional code \name introduces is negligible, we expect that the resulting additional execution time is also small.
In such a case, an external profiling tool might introduce too much noise and might be too coarse-grained in the time measurement to give accurate results.
We therefore choose to instrument the source code of the subject apps and add time-measurement code into them to have a fine-grained analysis of each module's overhead. 
We insert nanosecond-level \texttt{System.nanoTime()} timing functions just before and after the code block of each module, the sum of the readings of which will be the overall overhead of \name.

\textbf{Experimental setup.}
We use a Nexus 4 phone to evaluate the performance overhead.
As some modules of \name may introduce different overhead for Intent-based components and \texttt{Provider} (e.g., obtaining caller identity), we include both types of components for our evaluation.
More specifically, we use Terminal Emulator's \texttt{RemoteInterface} and Zirco Browser's \texttt{ZircoBookmarksProvider} as the evaluation subjects.
In particular, we consider the overhead of popping up alert dialogs as part of \texttt{SecActivity}'s overhead, because it delays the presence of the original callee Activity whereas the other three types of background components have no such concern.
Moreover, protecting Activity components generally uses the \texttt{alert} policies. 
For each app, we test 20 times and report the average for normal IPC latency and the execution time of each module in \name.
Since \texttt{startActivity()} is an asynchronous API, the normal IPC latency of an \activity is defined as the latency from the time when \texttt{startActivity()} is called to the time when the user interface is just to be displayed.
For \provider, since \texttt{ContentResolver.query()} is a synchronous API, we simply measure its execution time.

\textbf{Result overview.}
Table~\ref{tab:overhead} shows the performance results for the tested \activity and \provider, respectively.
We can see that the cumulative overhead (i.e., the worst-scenario overhead) is below 5\% for both components, with 4.42\% for \activity and 3.71\% for \provider.
Also, the absolute cumulative timing overhead is only 20.55ms and 0.4ms, which is unnoticeable to human users.
Moreover, we would like to underline that \name brings overheads only at the entry points of risky components, while existing defenses cause slowdown to the entire app or system.

\textbf{Overhead breakdown.}
We further perform a breakdown analysis of \name's overheads in each module.
As shown in Table~\ref{tab:overhead}, there are five modules corresponding to the following four scenarios (S1 to S4):

\begin{compactitem}
\item[S1]
\textit{Overhead on intra-app communication}: $t_1$ or $t_1{'}$. 
To normal intra-app communication, \name only introduces the binder analysis overhead $t_1$ or the overhead of getting the caller identity in \provider $t_1{'}$.
Both are small at 1.88\% and 2.22\%, respectively.

\item[S2]
\textit{Overhead on SQL injection}: $t_1{'} + t_{1.5}{'}$.
For \provider components, \name needs to filter malicious SQL strings, but such overhead is negligible (0.001ms only).

\item[S3]
\textit{Overhead on attacks that require manifest analysis and policy assessment}: $S1 + t_2/t_2{'} + t_3/t_3{'}$.
\name performs manifest analysis and policy assessment to stop hijacking attacks.
The additional delay introduced is also small at 0.29ms for $t_2 + t_3$ and 0.16ms for $t_2{'} + t_3{'}$.

\item[S4]
\textit{Overhead on attacks that trigger alert dialogs}: $S1 + S3 + t_4$.
Finally, in the worst scenario, \name introduces $\sim$10ms additional delay (i.e., $t_4$) to trigger the alert.
This delays the presence of the callee Activity.
\end{compactitem}

\begin{table}[t!]
\centering
\caption{Breakdown of \name's overheads.}
\vspace{-3ex}
\label{tab:overhead}
\scalebox{0.9}{
\begin{threeparttable}
\begin{tabu}{|c|c|c|c|}

\hline
\rowfont{\bfseries}
\rowcolor{LightGray}
Scenario & Category           & Time cost   & Overhead \% \\ 

\hline
\hline

\multicolumn{4}{|c|}{\textbf{Activity}} \\ \hline
\begin{tabular}[c]{@{}c@{}}Original \\ scenario\end{tabular} & Normal IPC latency: $t_0$ & 464.40ms & - \\ \hline
\multirow{5}{*}{\begin{tabular}[c]{@{}c@{}}Overheads \\ introduced \\ by \name\end{tabular}} & Binder analysis: $t_1$ & 8.73ms      & 1.88\%      \\ 
& Manifest analysis: $t_2$  & 0.24ms\tnote{\dag} & 0.05\%\tnote{\dag}  \\ 
& Policy assessment: $t_3$  & 0.05ms      & 0.01\%      \\ 
& Popping up alerts: $t_4$  & 11.53ms     & 2.48\%      \\ \cline{2-4} 
& Sum (worst-scenario)      & 20.55ms     & 4.42\%      \\

\hline
\hline

\multicolumn{4}{|c|}{\textbf{Provider}} \\ \hline
\begin{tabular}[c]{@{}c@{}}Original \\ scenario\end{tabular} & Normal IPC latency: $t_0{'}$ & 10.82ms    & -           \\ \hline
\multirow{5}{*}{\begin{tabular}[c]{@{}c@{}}Overheads \\ introduced \\ by \name\end{tabular}} & Getting caller identity: $t_1{'}$ & 0.24ms      & 2.22\%      \\ 
& SQL filtering: $t_{1.5}{'}$  & 0.001ms     & 0.01\%      \\ 
& Manifest analysis: $t_2{'}$  & 0.146ms\tnote{\dag} & 1.35\%\tnote{\dag}  \\ 
& Policy assessment: $t_3{'}$  & 0.014ms      & 0.13\%      \\ \cline{2-4} 
& Sum (worst-scenario)         & 0.401ms     & 3.71\%      \\ \hline

\end{tabu}
\begin{tablenotes}
\item[\dag] \name analyzes manifest only once for the entire lifecycle of the app.
  In the \activity context, manifest analysis takes 2.41ms in the first run and zero for the rest of runs.
  Similarly, the analysis of the first run on \provider takes 1.46ms.
  Therefore, we calculate an estimated value by assuming that there are ten IPC transactions in a lifecycle of the app. 
\end{tablenotes}
\end{threeparttable}
}
\end{table}

\section{Discussion}
\label{sec:discuss}

We now discuss \name's several limitations and potential improvements to be made.
Firstly, requiring a re-compilation of the Android app (not a re-design of the software) is a limitation of \name. However, it is not only targeted to new apps but also existing ones, since Android apps typically require frequent updates (Google Play enforces mandatory app updates by default) and re-compilation anyway.
Secondly, as we have discussed in \mysec\ref{sec:policy}, \name currently cannot handle hijacking via one or multiple middle component(s) in the callee app. 
But for the middle-app problem, since \name only needs to differentiate an external or an internal IPC in this scenario, in principle \name can handle it.
The only exception is that for \texttt{alert} policies, an attack app may leverage the middle app's identity to trick users into clicking the ``allow'' button.
Thirdly, although \name has minimized usability issues for most user operations by trusting intra-app IPC (\mysec\ref{sec:policy}), its \texttt{alert} policies may still cause a few usability issues.
These could be further mitigated by the fact that \name only needs to enforce policies on risky components (\mysec\ref{sec:other}).
Since the percentage of risky components is small (see Table~\ref{tab:dataset}), we expect the usability issues to be minimal.
Moreover, we can leverage the recent advance of user-driven access control~\cite{AUDACIOUS16} and knowledge in the HCI field to further minimize its impact.
Finally, \name cannot protect third-party components of which developers do not own source code (e.g., advertisement libraries);
but library providers can periodically push updates to enhance the security of their components.



\section{Conclusion and Future Works}
\label{sec:conclude}

In this paper, we presented a practical and lightweight approach called \name to defend against component hijacking in Android apps.
\name is essentially a secure component library that performs in-app mandatory access control on behalf of app components.
We designed six mandatory policies for \name to stop attacks originated from both system weaknesses and common developer mistakes.
We have implemented a proof-of-concept \name prototype and demonstrated its efficacy and efficiency.
In the future, we will try to integrate \name into Boxify~\cite{Boxify15} after its code is released.

\section*{Acknowledgements}
We thank all the reviewers of this paper for their valuable comments.
This work is partially supported by the Singapore National Research
Foundation under NCR Award Number NRF2014NCR-NCR001-012.

\bibliographystyle{ACM-Reference-Format}
\bibliography{main}

\end{document}